\documentstyle[epsf]{mn}
\newif\ifAMStwofonts

\def\eps@scaling{.95}
\def\epsscale#1{\gdef\eps@scaling{#1}}

\def\plotone#1{\centering \leavevmode
    \epsfxsize=\eps@scaling\columnwidth \epsfbox{#1}}

\def\url#1{{\ttfamily\def\/{/\discretionary{}{}{}}#1}}

\def\kpch{\mbox{$h^{-1}$kpc}}

\def\Mpch{\mbox{$h^{-1}$Mpc}}

\def\msunh{\mbox{$h^{-1}$M$_\odot$}}
\def\Msunh{\mbox{$h^{-1}$M$_\odot$}}

\def\LCDM{{\char'3CDM}}

\def\mathnew{\mathsurround=0pt}
\def\simov#1#2{\lower .5pt\vbox{\baselineskip0pt
    \lineskip-.5pt\ialign{$\mathnew#1\hfil##\hfil$\crcr#2\crcr\sim\crcr}}}

\def\'#1{\ifx#1i{\accent"13\i}\else{\accent"13#1}\fi}

\ifoldfss
  \ifCUPmtlplainloaded \else
    \NewTextAlphabet{textbfit} {cmbxti10} {}
    \NewTextAlphabet{textbfss} {cmssbx10} {}
    \NewMathAlphabet{mathbfit} {cmbxti10} {} 
    \NewMathAlphabet{mathbfss} {cmssbx10} {} 
  \fi
  \ifAMStwofonts
    \ifCUPmtlplainloaded \else
      \NewSymbolFont{upmath} {eurm10}
      \NewSymbolFont{AMSa} {msam10}
      \NewMathSymbol{\upi}     {0}{upmath}{19}
      \NewMathSymbol{\umu}     {0}{upmath}{16}
      \NewMathSymbol{\upartial}{0}{upmath}{40}
      \NewMathSymbol{\leqslant}{3}{AMSa}{36}
      \NewMathSymbol{\geqslant}{3}{AMSa}{3E}

      \let\leq=\leqslant 
      \let\geq=\geqslant 
    \fi
  \fi
\fi 

\ifnfssone
  \newmathalphabet{\mathit}
  \addtoversion{normal}{\mathit}{cmr}{m}{it}
  \addtoversion{bold}{\mathit}{cmr}{bx}{it}
  \newmathalphabet{\mathbfit} 
  \addtoversion{normal}{\mathbfit}{cmr}{bx}{it}
  \addtoversion{bold}{\mathbfit}{cmr}{bx}{it}
  \newmathalphabet{\mathbfss} 
  \addtoversion{normal}{\mathbfss}{cmss}{bx}{n}
  \addtoversion{bold}{\mathbfss}{cmss}{bx}{n}
  \ifAMStwofonts
    \ifCUPmtlplainloaded \else
      \UseAMStwoboldmath
      \makeatletter
      \new@mathgroup\upmath@group
      \define@mathgroup\mv@normal\upmath@group{eur}{m}{n}
      \define@mathgroup\mv@bold\upmath@group{eur}{b}{n}
      \edef\UPM{\hexnumber\upmath@group}
      \new@mathgroup\amsa@group
      \define@mathgroup\mv@normal\amsa@group{msa}{m}{n}
      \define@mathgroup\mv@bold\amsa@group{msa}{m}{n}
      \edef\AMSa{\hexnumber\amsa@group}
      \makeatother
      \mathchardef\upi="0\UPM19
      \mathchardef\umu="0\UPM16
      \mathchardef\upartial="0\UPM40
      \mathchardef\leqslant="3\AMSa36
      \mathchardef\geqslant="3\AMSa3E

      \let\leq=\leqslant 
      \let\geq=\geqslant 
    \fi
  \fi
\fi 

\ifnfsstwo
  \DeclareMathAlphabet{\mathbfit}{OT1}{cmr}{bx}{it}
  \SetMathAlphabet\mathbfit{bold}{OT1}{cmr}{bx}{it}
  \DeclareMathAlphabet{\mathbfss}{OT1}{cmss}{bx}{n}
  \SetMathAlphabet\mathbfss{bold}{OT1}{cmss}{bx}{n}
  \ifAMStwofonts
    \ifCUPmtlplainloaded \else
      \DeclareSymbolFont{UPM}{U}{eur}{m}{n}
      \SetSymbolFont{UPM}{bold}{U}{eur}{b}{n}
      \DeclareSymbolFont{AMSa}{U}{msa}{m}{n}
      \DeclareMathSymbol{\upi}{0}{UPM}{"19}
      \DeclareMathSymbol{\umu}{0}{UPM}{"16}
      \DeclareMathSymbol{\upartial}{0}{UPM}{"40}
      \DeclareMathSymbol{\leqslant}{3}{AMSa}{"36}
      \DeclareMathSymbol{\geqslant}{3}{AMSa}{"3E}

      \let\leq=\leqslant 
      \let\geq=\geqslant 
    \fi
  \fi
\fi 

\ifCUPmtlplainloaded \else
  \ifAMStwofonts \else 
    \def\upi{\pi}
    \def\umu{\mu}
    \def\upartial{\partial}
  \fi
\fi

\title[Strong Lensing and Dark Energy Models]{Strong Gravitational Lensing and 
Dynamical Dark Energy}

\author[A.V. Macci\`o] {Andrea V. Macci\`o \\ Institute for Theoretical
  Physics, University of Z$\ddot u$rich,
  CH-8057 Z$\ddot u$rich, Switzerland \\ Physics Department G. Occhialini, 
Universit\`a degli Studi di Milano-Bicocca, Piazza della Scienza 3, I-20126
  Milan, Italy \\ INFN, via Celoria 16, I-20133 Milan, Italy}

\date{Draft version \today}
\pagerange{\pageref{firstpage}--\pageref{lastpage}}
\pubyear{2002}

\begin{document}

\maketitle

\label{firstpage}

\begin{abstract}

We study the strong gravitational lensing properties of galaxy clusters
obtained from N-body simulations with different kind of Dark Energy (DE).
We consider both dynamical DE, due to a scalar field self--interacting through 
Ratra--Peebles (RP) or SUGRA potentials, and DE with constant negative $w=p/\rho=
-1$ (\LCDM). We have 12 high resolution lensing systems for each cosmological
model with a mass greater than $5.0 \times 10^{14}$\Msunh. 
Using a Ray Shooting technique we make a detailed
analysis of the lensing properties of these clusters with particular attention
to the number of arcs and their properties (magnification, length and width).
We found that the number of giant arcs produced by galaxy clusters changes
in a considerable way from \LCDM~ models to Dynamical Dark Energy models 
with a RP or SUGRA potentials. These differences originate from the different epochs of
cluster formation and from the non-linearity of the strong lensing effect.
We suggest the Strong lensing is one of the best tool to discriminate among
different kind of Dark Energy.

\end{abstract}
\begin{keywords}
methods: analytical --- methods: numerical --- galaxies:
clusters: general --- cosmology: theory --- dark matter --- galaxies:
halos
\end{keywords}

\section{Introduction}

The mounting observational evidence for the existence of Dark Energy (DE), 
which probably accounts for $\sim 70\%$ of the critical density of the
Universe 
\cite{Perlmutter,Riess,Tegmark01,Netterfield,Pogosian03,Efstathiou2,Percival,spergel2003}, 
rises a number of questions
concerning galaxy formation. The nature of DE is suitably described by
the parameter $w=p/\rho$, which characterizes its equation of state.
The \LCDM~model ($w=-1$) was extensively studied during the last
decade. 
Recently much more attention was given to physically
motivated models with variable $w$ \cite{Mainini03a}, for which
a number of N--body simulations have been performed (Klypin et al 2003,
KMMB03 hereafter, Dolag et al. 2003, Linder \& Jenkins 2003, Macci\`o et al. 2004).  
One of the main results of KMMB03 was that dynamical DE halos are 
denser than those with the standard \LCDM~ one. 
In this work we want to analyze the impact of this higher concentration on the
strong lensing properties of the cluster size halos.

Was first noted by Bartelmann et al. (1998) (for OCDM, SCDM and \LCDM~ cosmology)
that the predicted number of giant arcs varies by orders of magnitude among
different cosmological models. The agreement between data and \LCDM~
simulation was tested by many authors (see Meneghetti et
al 2000, Dalal et al. 2003, Wambsganss et al. 2004) but the situation is still unclear. 
A direct comparison of arcs statistic with observational data is out of the
scope of this work, what we want to point out is the capability of Strong
Lensing to discriminate between different kinds of Dark Energy 
(a similar paper but for a different choice of the dynamical DE parameters 
was recently submitted by Meneghetti et al. (2004)).

Here, using a Ray Shooting technique, we make a lensing analysis of dark matter halos
extracted from N-body simulations of cosmological models with varying $w$ 
arising from physically motivated potentials 
which admit tracker solutions. In particular, we focus on the two 
most popular variants of dynamical DE \cite{wett1,RP,wett2}.  
The first model was proposed by Ratra \& Peebles (1984, RP hereafter) and it yields
a rather slow evolution of $w$. The second model 
\cite{BraxMartin99,BraxMartinRiazuelo,BraxMartin00} is based on potentials found in 
supergravity (SUGRA) and it results in a much faster evolving $w$.
Hence, RP and SUGRA potentials cover a large spectrum of evolving
$w$. These potentials are written as
\begin{eqnarray}
V(\phi) &=& \frac{\Lambda^{4+\alpha}} {\phi^\alpha} \qquad RP, \\
V(\phi) &=& \frac{\Lambda^{4+\alpha}}{\phi^\alpha} \exp (4\pi G \phi^2)~~~ SUGRA.
\end{eqnarray}
Here $\Lambda$ is an energy scale, currently set in the range
$10^2$--$10^{10}\, $GeV, relevant for the physics of fundamental
interactions. The potentials depend also on the exponent $\alpha$.
The parameters $\Lambda$ and $\alpha$ define the DE density parameter
$\Omega_{DE}$. However, we prefer to use $\Lambda$ and $\Omega_{DE}$
as independent parameters.  Figure~10 in Mainini et al. (2003b) gives
examples of $w$ evolution for RP and SUGRA models. 

The SUGRA model considered in this paper has $\Lambda=10^3$ GeV this implies 
$w=-0.85$ at $z=0$, but $w$ drastically changes with redshift: $w\approx -0.4$ at $z=5$.
The first RP model (RP$_1$) has the same value for $\Lambda$ of the SUGRA model. 
At redshift $z=0$ it has $w=-0.5$; then value of $w$ gradually changes with the 
redshift: at $z=5$ it is close to $-0.4$.  
Although the $w$ interval spanned by this RP model covers values
significantly above -0.8 (not favored by observations), this case is
still important both as a limiting reference case and to emphasize
that models with constant $w$ and models with variable $w$ produce
different results even if average values of $w$ are not so different.
For the second RP model (RP$_2$) we have chosen $\Lambda=10^{-8}$ GeV, in this case the
value of the state parameter at redshift $z=0$ is the same of SUGRA:
$w(z=0,\Lambda=10^{-8}$ GeV$)=-0.85$. This model is certainly better in agreement with
CMB constrains but it loses most of its interest from a theoretical point of
view: such a small value of $\Lambda$ has not any clear connection with the
physics of fundamental interactions and so it has exactly the same ``fine
tuning'' problem of the \LCDM~ model. 

We have normalized all the models in order to have today the same value of the $rms$ density
fluctuation on a scale of 8 \Mpch~, that has been chosen as $\sigma_8=0.8$.

\section{N-body Simulations}
The Adaptive Refinement  Tree code (ART;  Kravtsov, Klypin \& Khokhlov 
1997) was used   to run the  simulations.  The ART code starts  with a
uniform  grid, which  covers the whole   computational box. This  grid
defines  the  lowest (zeroth) level of   resolution of the simulation.
The standard Particles-Mesh algorithms are used to compute density and
gravitational  potential  on the   zeroth-level  mesh.  The  ART  code
reaches   high force resolution by refining   all high density regions
using   an  automated  refinement   algorithm.   The   refinements are
recursive:  the refined regions can  also be  refined, each subsequent
refinement having half of the previous level's cell size. This creates
a hierarchy of refinement  meshes  of different resolution, size,  and
geometry  covering regions of  interest. Because each individual cubic
cell can be refined, the shape of the refinement mesh can be arbitrary
and match effectively the geometry of the region of interest.

The criterion for refinement is the local density of particles: if the
number of particles in a mesh cell (as estimated by the Cloud-In-Cell
method) exceeds the level $n_{\rm thresh}$, the cell is split
(``refined'') into 8 cells of the next refinement level.  The
refinement threshold depends on the refinement level. For the zero's
level it is $n_{\rm thresh}=2$. For the higher levels it is set to
$n_{\rm thresh}=4$. The code uses the expansion parameter $a$ as the
time variable.  During the integration, spatial refinement is
accompanied by temporal refinement.  Namely, each level of refinement,
$l$, is integrated with its own time step $\Delta a_l=\Delta a_0/2^l$,
where $\Delta a_0= 3\times 10^{-3}$ is the global time step of the zeroth refinement
level.  This variable time stepping is very important for accuracy of
the results.  As the force resolution increases, more steps are needed
to integrate the trajectories accurately.  Extensive tests of the code
and comparisons with other numerical $N$-body codes can be found in
Kravtsov (1999) and Knebe et al. (2000).  The code was modified to
handle DE of different types (Mainini et al 2003b \& KMMB03).

We performed a low resolution simulation for each model with the following parameters:
box size: 320 \Mpch, number of particles: $128^3$, force resolution: 9.2
\kpch, all the simulations have the same initial random seed so at $z=0$ the
clusters are more or less in the same positions.
Than we selected the four massive clusters in the \LCDM~ simulation
and re-run them with a mass resolution 64 times higher. The same clusters are
also re-run with the same resolution also in the RP and SUGRA models.
At the end we have 12 lensing systems (each cluster can be
seen by three different orthogonal directions) for each cosmological model,
with a mass resolution of $2.03 \times 10^{10} \msunh$ and a force resolution
of 4.8 \kpch. A complete list of simulation parameters is contained in table~1.

\begin{table}
\centering
\begin{minipage}{140mm}

\begin{tabular}{llllll} 
\hline
\hline
{Model} & {$\Lambda$} 
& {Box} & {Np} & {M$_{res}$.} & {F$_{res}$.} \\

& (GeV)&  (\Mpch) & & (\Msunh) & (\kpch) \\

\hline
\hline
RP$_1$   & 10$^3$ & 320 & 512$^3$ & $2.03\times 10^{10}$ & 4.8 \\ 
RP$_2$  & 10$^{-8}$ & 320 & 512$^3$ & $2.03\times 10^{10}$ & 4.8 \\ 
SUGRA & 10$^3$  & 320 & 512$^3$ & $2.03\times 10^{10}$ & 4.8 \\ 
\LCDM & 0 & 320 & 512$^3$ & $2.03\times 10^{10}$ & 4.8 \\ 

\hline\hline 

\end{tabular} 
\end{minipage}
\caption{Parameters of simulations}
\end{table} 


\section{Lensing Simulations}

In order to compute arc statistics for the 
models discussed above,
we adopted a technique similar to
the one originally proposed by Bartelmann \& Weiss (1994).
We center the cluster in a cube of 4 \Mpch~side length
and study three lenses, obtained by projecting the particle positions along
the coordinate axes. 
This grants us a total of 12 lens planes per model that we 
treat as though being due to independent clusters, for our present purposes.

\begin{figure}
\plotone{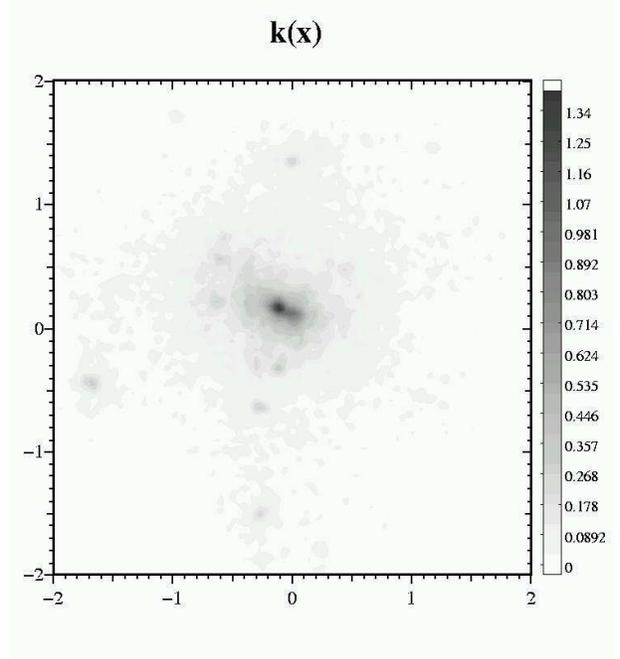}
\caption{\small Convergence map for one of the \LCDM~ clusters. The side length
  is 4 \Mpch.}
\label{fig:conv}
\end{figure}

We then divide the projected density field $\Sigma$ by the critical 
surface mass density for lensing
\begin{equation}
\Sigma_{cr} = { c^2 \over {4 \pi G}} { D_S \over {D_L D_{LS}}} ~,
\end{equation}
so  obtaining the convergence $k$. Here $c$ is the speed of light, 
$G$ is the gravitational constant, while
$D_L$, $D_S$, $D_{LS}$ are the angular-diameter distances between lens and
observer, source and observer, lens and source, respectively.
Once we set the lens and source redshift, the value of the
angular diameter distance depends on the cosmological model.
We detail this point in the next section.
In Figure \ref{fig:conv} we show
the convergence map for one of the cluster, 
whose length scale size is 4 \Mpch. 
The deflection angle due to this 2D particle distribution, on a given point
$\vec x$ on the lens plane reads:
\begin{equation}
\vec \alpha(\vec x) = \sum_{j=1}^N {{4 G } \over 
{c^2}} { {m_j} \over { \vert \vec x - \vec y_j \vert}} ~.
\label{eq:alpha}
\end{equation}
Here $\vec y_j$ is the position of the $j$-th particles and $N$ is the total
number of particles.

As direct summation requires a long time, we sped up the code by using a
P$^3$M--like algorithm: the lens plane was divided into 128$\times$128 cells and
direct summation was applied to particles belonging to the same cell of $\vec
x$ and for its 8 neighbor cells. Particles in other cells were then seen as a
single particle in the cell baricenter, given the total mass of the particles inside the cell.

The deflection angle diverges when the distance between a light ray and a
particle is zero. To avoid this unwanted feature we introduce a softening parameter
$\epsilon$ in eq:(\ref{eq:alpha}); the value 
$\epsilon$ is tuned on the resolution of the current simulation.

We start to compute $\vec \alpha (\vec x)$ on a regular grid of 
256$\times$256 test rays that covered the central quarter of the lens plane, 
then we propagate a bundle of 2048$\times$2048 light rays and determine the
deflection angle on each light ray by bicubic interpolation
amongst the four nearest test rays (see section \ref{sec:AA} for further
discussion on the effects of the adopted resolution in the lens mapping).

The relation between images and sources position is given by the lens
equation:
\begin{equation}
\vec y = \vec x - \vec \alpha (\vec x) 
\label{eq:len}
\end{equation}
and the local properties 
of the lens mapping are then described by the Jacobian matrix
of the lens equation,
\begin{equation}
A_{hk}(\vec x ) = { {\partial y_h } \over {\partial x_k}} = \delta_{hk} - 
{ {\partial \alpha_h \over \partial x_k}}
\end{equation}
The shear components $\gamma_1$ and $\gamma_2$ are found from $A_{hk}$ through
the standard relations:
\begin{equation}
\gamma_1(\vec x) = -{1\over 2} [ A_{11}(\vec x ) - A_{22}(\vec x)] , 
\end{equation}
\begin{equation}
\gamma_2(\vec x) = -{1\over 2} [ A_{12}(\vec x ) + A_{21}(\vec x)] , 
\end{equation}
and the magnifications factor $\mu$ is given by the Jacobian determinant,
\begin{equation}
\mu(\vec x)  = {1 \over { \det A}} = [ A_{11}(\vec x)A_{22}(\vec x)- 
A_{12}(\vec x) A_{21}(\vec x)]^{-1} .
\end{equation}

Finally, the Jacobian determines the location of the critical curves $\vec x_c$
on the lens plane, which are defined by $\det A(\vec x_c) = 0$. Because of the
finite grid resolution, we can only approximately locate them by looking for
pairs of adjacent cells with opposite signs of $\det A$. Then, the lens 
equations
\begin{equation}
\vec y_c = \vec x_c - \vec \alpha(\vec x_c),
\label{eq:lens}
\end{equation}
yield the corresponding caustics $\vec y_c$, on the source plane.

\subsection{Sources deformation}

For statistical purposes one has to distribute and map a large number of
sources. We are interested in arc properties and arcs form near caustics; so for
numerical efficiency we have to distribute less sources in those part of the
source plane that are far away from any caustics, and more sources close or
inside the caustics. We follow the method introduced by Miralda-Escud\`e
(1993) and later adapted to non-analytical models by Bartelmann \& Weiss
(1994). In the previous section we have obtained the deflection angles for the 
$\vec x_{ij}$ (with i,j=1,...,2048) positions on the lens (or image) plane, 
using the lens equation (\ref{eq:lens}) we can obtain the corresponding
positions on the source plane $\vec y_{ij}(\vec x_{ij})$, as usual we call
this discrete transformation the {\it mapping table}.

We model elliptical sources with axial ratios randomly drawn from the interval
$[0.5;1]$ and area equal to that of a circle with radius $r_s = 0.5''$.
We first distribute sources on a coarse grid of 32x32, defined in the
central quarter of the source plane covered by the light rays traced (due to
convergence only a restricted part of the source plane can be reached by the
light rays traced from the observer through the lens plane). From the mapping
table we have obtained the magnification ($\mu$), if it changes by more than
one (absolute value) between two sources, we place an additional source
between both, in this way we increase the resolution by a factor 2 in each 
dimension. For the n-th iteration of source positions the criterium to add
additional sources is that magnification changes by $2^{n-1}$. We repeat this
procedure four times to obtain the final list of source positions.
To compensate for this artificial increase in the source number density we
assign a statistical weight of $2^{1-n}$ to each image of a source placed
during the $n$-th grid refinement. On average we have about 15000 sources for
each lensing system.

\subsection{Arcs Analysis}
\label{sec:AA}

To find the images of an extended source, all images-plane positions 
$\vec x$ are checked if the corresponding entry in the map table 
$\vec y$ lies within the source: i.e. for an elliptical source with axes
$a,b$ and centered in $(y^c_1;y^c_2)$ it is checked if:
\begin{equation}
{ {(y_1-y_1^c)^2} \over a^2} + {{(y_2-y_2^c)^2} \over b^2} \leq 1,
\end{equation}
where $(y_1,y_2)$ are the components of the vector $\vec y$.

Those points fulfilling the previous equation are part of one of the source
images and are called image points. We then use a standard {\it friends-of-friends}
algorithm to group together image points within connected regions, since they
belong to the same image (the number of images of one source ranges from 1 to
5 for our clusters).

We measure arc properties using a method based on Bartelmann \& Weiss (1994).
The arc area and magnification are found by summing the areas of the pixel
falling into the image.
Arc lengths are estimated by first finding the arc center, then finding the
arc pixel farest from the centroid as well as the pixel farest from this
pixel. The arc length is then given by the sum of the lengths of the two lines
connecting these three points. The arc width is defined as the ratio between
the arc area and the arc length.

\begin{figure}
\plotone{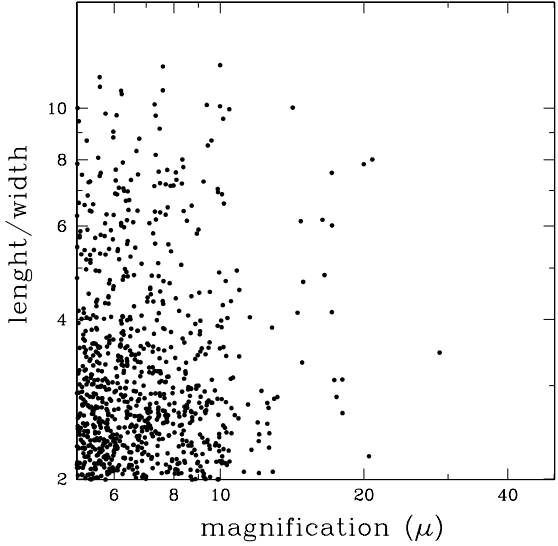}
\caption{\small Magnification vs. length/width ratio for \LCDM~ clusters. These
  two quantities are equal for an isothermal sphere lens.
}
\label{fig:lw}
\end{figure}

In Figure \ref{fig:lw} we plot the relation between length/width ratio and
magnification; we found a good agreement with previous results obtained by
Dalal et al. (2003). The scatter in this relation is due to local fluctuations in
the surface mass density, highly distorted images are also highly magnified,
but the converse is not always true.

\begin{figure}
\plotone{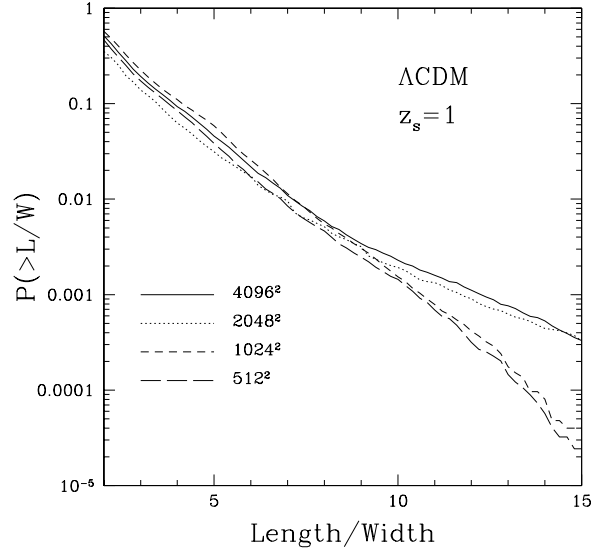}
\caption{\small Fraction of sources mapped in an arc vs length/width ratio of
  the arc for different values of the resolution of the lensing mapping.
  All curves are for the  \LCDM~ model with $z_l=1.0$ and $z_s=0.3$.
}
\label{fig:res}
\end{figure}

Before proceeding with our analysis we have performed some tests on the resolution
adopted in our ray shooting code; figure \ref{fig:res} shows the fraction of
sources (number of sources divided by the total number) vs. their length/width
for different values of the resolution of the lens mapping grid $N_{hr}^2$ (results are
for the \LCDM~model with $z_l=0.3$ and $z_s=1.0$). 
If the resolution of the lens mapping is not high enough the critical curves
are too small compared to the source size and a spurious cut off in the number
of arcs with $L/W> 10$ appears. This cut off is totally artificial and it
vanishes for $N_{hr} \geq 2048$. As figure \ref{fig:res} shows the results
are stable also for a higher value of $N_{hr}$ (4096), then in order to have a
good compromise between resolution and computational time we have 
adopted $N_{hr}= 2048$ in the following.

\section{Arcs Statistics }

In this paper we aim to compare the lensing properties of 
a given cluster as it appears in different cosmological 
models. There are three main features that affect the number of giant arcs:
the concentration of the halo, the total number of lensing systems at a given
redshift and the value of the critical surface mass density ($\Sigma_{cr}$).

\begin{figure}
\plotone{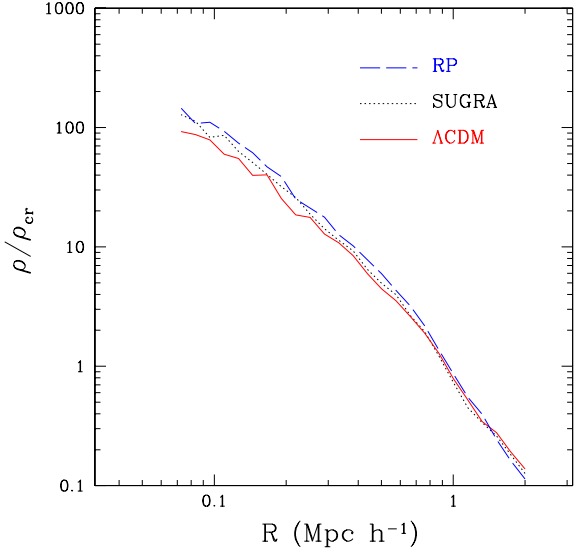}
\caption{\small Density profile of the same halo simulated in different
  model. The red curve is for \LCDM~, the black one for SUGRA and the blue for RP$_1$.
The halo has a virial mass of $6 \times 10^{14} \msunh~$.
}
\label{fig:prof}
\end{figure}

As predicted analytically by Bartelmann et al (2002) (for constant $w$ models)
and first noted in numerical Nbody simulations by KMMB03, 
and then confirmed by Dolag et al (2003) and Linder \& Jenkins (2003), 
the concentration of dynamical DE halos is greater than the concentration of \LCDM~ ones. 
Here we use the same definition of concentration of KMMB03: the ratio of
the radius at the overdensity of the \LCDM~ model (103 times the critical
density) to the characteristic (``core'') radius of the NFW profile.
(see however KMMB03 for more details).
A greater concentration increases the probability of forming giant arcs.
In Figure \ref{fig:prof} we report the density profile of the same halo 
simulated in different cosmological models. The RP$_1$ halo is clearly denser and
more concentrated than the \LCDM~ halo with the SUGRA halo 
laying in between; the RP$_2$ halo (which is not shown in this plot) has a
concentration parameter close to the one of the \LCDM~ model.

\begin{figure}
\plotone{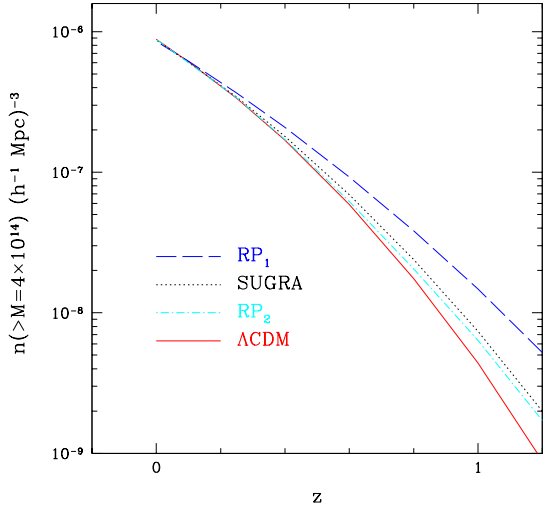}
\caption{\small Number density of halos with a mass greater than $4.0\times
  10^{14} \msunh$ for \LCDM~ (low solid curve), SUGRA (dotted curve) and the
  two RP models. The value of $\Lambda$ is $10^3$ GeV for both RP$_1$ and
  SUGRA and $10^{-8}$ GeV for RP$_2$.
}
\label{fig:mf}
\end{figure}

The expected number of objects with a mass 
exceeding $4\times 10^{14} \msunh$
(in order to produce multiple images) at a given redshift (in this case
$z=0.4$) can be estimated using a Press \& Schechter 
formalism (see Mainini et al 2003a). In dynamical DE, objects form earlier
than in \LCDM, so we have more lensing systems per Mpc$^3$ at
$z=0.3$. 
This can be taken into account by multiplying the number of arcs by 1.3, 1.21
and 1.12 in RP$_1$, SUGRA and RP$_2$ respectively.
(In Figure \ref{fig:mf} we report the evolution with redshift of the mass
function for a mass threshold of $4.0\times 10^{14} \msunh$).

\begin{figure}
\plotone{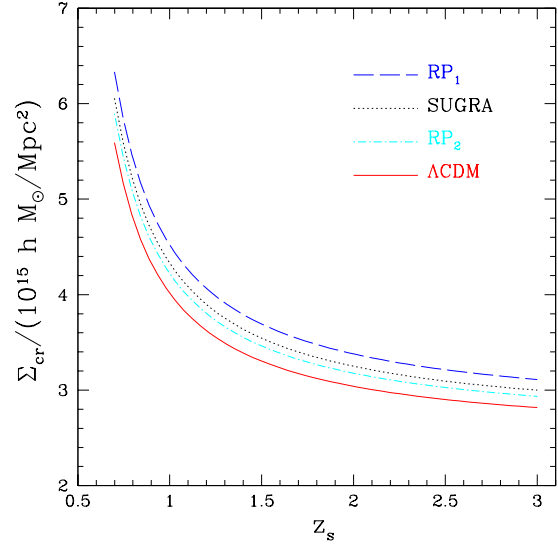}
\caption{\small Value of the critical surface mass density $\Sigma_{cr}$ in
  different cosmological models, for all the curves we choose a lens redshift
  $z_L=0.4$
}
\label{fig:sigma}
\end{figure}

The evolution of the scale factor $a$ with time also depends on the model.
This implies that, at a given redshift $z=1/a -1$, the
angular diameter distance $D_{ad}$ is 
model dependent;
in fact its value is given by:
\begin{equation}
D_{ad}(a) = {a c \over {H_0}} \int_a^1 \sqrt{ {a (1-\Omega_{DE}(a))} \over
  {\Omega_{m,0}}} ~~\rm{d} a ~.
\label{eq:da}
\end{equation}
Here $c$ is the speed of light, $H_0$ and $\Omega_{m,0}$ are the present value
of the Hubble constant and the matter density parameter and $\Omega_{DE}(a)$ 
gives the evolution of the DE density parameter with the expansion factor.
To compute $\Omega_{DE}(a)$ for RP and SUGRA models we have used the analytical
expression of Mainini et al (2003b).
In Figure \ref{fig:sigma} we show
the value of the critical surface mass
density for the adopted cosmological models. 
The different values for $\Sigma_{cr}$ mean that a \LCDM~ halo 
yields more arcs than a dynamical DE halo, if they have the same surface mass density.
The effect of the different values of the angular diameter distance tends therefore to
reduce the number of arcs in dynamical DE models.

\begin{figure}
\plotone{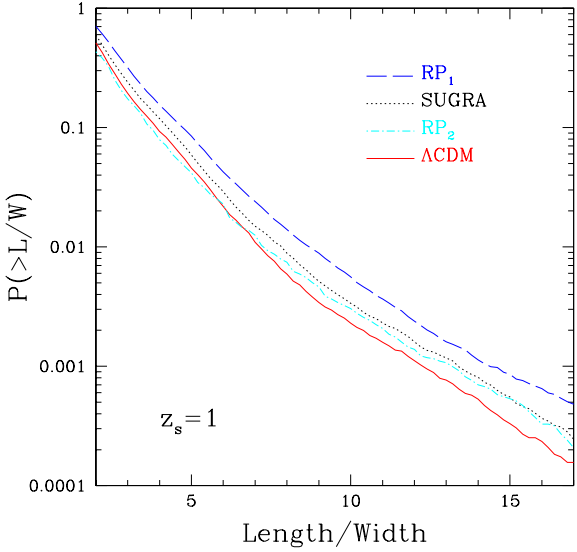}
\caption{\small Fraction of sources mapped in an arc vs length/width ratio of
  the arc. Upper curve is RP$_1$ (blue) than from top to bottom we have SUGRA
  (red), RP$_2$ (cyan) and \LCDM~ (black) results.
}
\label{fig:arcs}
\end{figure}

\begin{figure}
\plotone{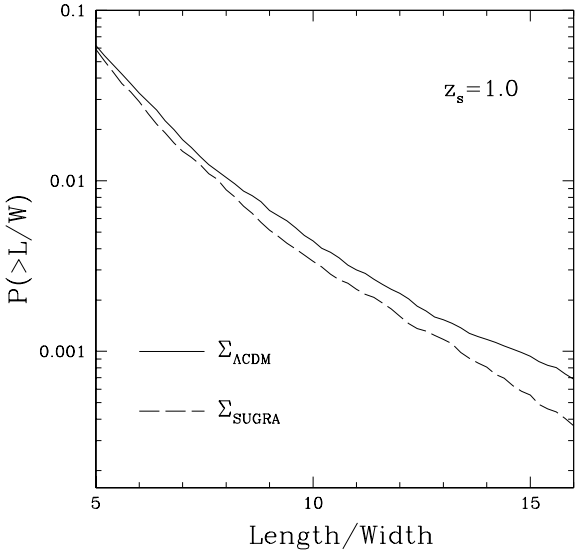}
\caption{\small Same plot of Figure \ref{fig:arcs} for the SUGRA model where
  the arc properties are computed using $\Sigma_{cr}(\LCDM~)$ (solid line) and
  $\Sigma_{cr}(SUGRA)$ (dashed line).
}
\label{fig:arcs2}
\end{figure}

The first result of our analysis is shown in 
Figure \ref{fig:arcs}, where
we plot the fraction of sources (number of sources divided by the total 
number) vs. their length/width ratio for a cluster (lens) redshift $z_l=0.3$, 
where lensing is most efficient for a source redshift of 1.0 (as shown later
in figure \ref{fig:arcsz}).
As expected, the RP$_1$ model produces more distorted images, due to its more
concentrated halos. The SUGRA and RP$_2$ models are quite similar for $L/W>10$ 
and they lay in between RP$_1$ and \LCDM~ which, as expected, produces less
highly distorted images than the dynamical DE models. We want to underline that part of
the higher lensing signal due to the higher concentration of dark matter halos
in such models is canceled by the increased $\Sigma_{cr}$ value. 
This effect is clearly illustrated in figure \ref{fig:arcs2} where we 
have computed the
arc statistics in a SUGRA model using the \LCDM~ critical surface density.
As expected, we have obtained a result for SUGRA that is closer to the RP$_1$ one.

\begin{figure}
\plotone{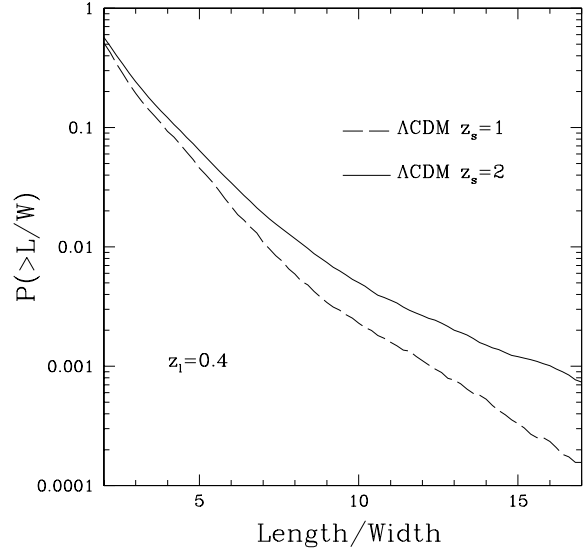}
\caption{\small Arcs counts for two different values of the sources redshift in the
  \LCDM~ model.
}
\label{fig:cfr}
\end{figure}

\begin{figure}
\plotone{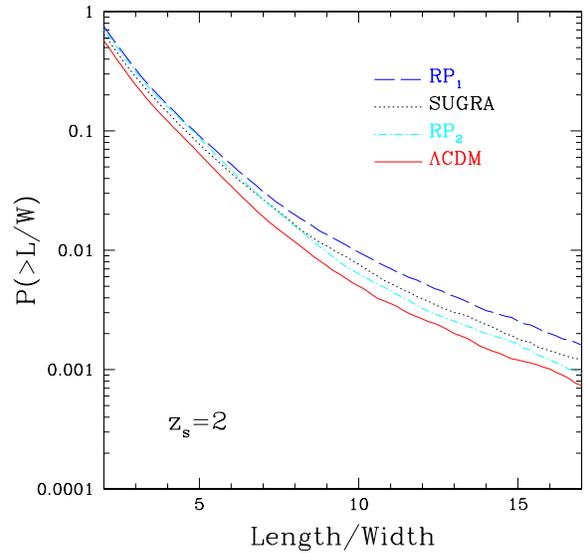} 
\caption{\small The same of Figure \ref{fig:arcs} for $z_s=2$. With this value
  for the source redshift the number of arcs increases in all the models.
}
\label{fig:arcs.2}
\end{figure}

\begin{figure}
\plotone{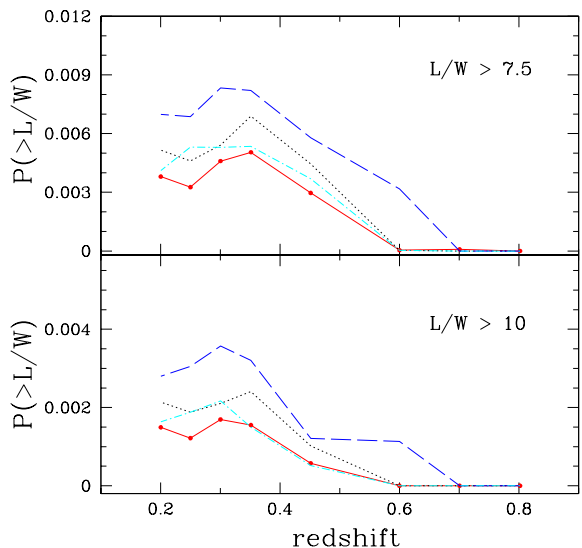}         
\caption{\small Number of arcs {\it vs} the lens redshift
  for two different thresholds of the $L/W$ ratio: 10 and 7.5. Solid line
  is for \LCDM~, dashed for RP$_1$, dot-dashed for RP$_2$  and dotted for SUGRA ($z_s=1.0$).}
\label{fig:arcsz}
\end{figure}

As pointed out by many authors (Wambsganss et al. 2003, Dalal et al. 2003) the
number of arcs that a cluster is able to produce is strongly related to
the redshift of the sources (although the strength of this effect is not 
yet completely understood, different authors found different results).
In Figure \ref{fig:cfr} we plot the arc number counts for the \LCDM~ model
for $z_s=1$ (dashed line) and $z_s=2$ (solid line). As in previous work we
found that the number of arcs increases if we increase the source redshifts.
Figure \ref{fig:arcs.2} shows the same results of Figure \ref{fig:arcs} but for
$z_s=2$.

As expected, the total number of arcs increases in all cosmological models.
Again we have a sort of hierarchy of results according to 
what is expected from the dynamical evolution of dark matter halos in the
corresponding cosmological model.

Moreover, due to the lower difference in the value of $\Sigma_{cr}$ for this
sources redshift, the four models are better separated, expecially the SUGRA 
and RP$_2$ ones. A difference between these two models is somewhat expected
even if they have the same value of state parameter today ($w=-0.84$), this
arises from the different evolution of $w$: in SUGRA it drastically
changes with redshift ($w=-0.4$ at $z=5$) when it is more constant in RP$_2$
($w=-0.72$ at $z=5$). 

As final result in figure \ref{fig:arcsz} we show the evolution with the
lens redshift of the number of arcs for two different choices of the
length/width ration: 10 and 7.5 (the redshift of the sources is $z_s=1.0$).
On average the the RP$_1$ model is always above the others, 
instead the different between SUGRA and \LCDM~ is more or less constant at all redshift
and the lensing signal decreases rapidly for $z>0.45$. 

The RP$_2$ model has a sort of double behavior: it is close to \LCDM~ for
$z>0.35$ but it is more similar to SUGRA for $z<0.35$, we think that this
bimodality is due again to the evolution of the state parameter in this model
expecially if compared to the SUGRA one: the ration $w_{RP_2}/w_{SU}$
decreases with redshift towards unity at $z=0$, so it is less different from
\LCDM~ at high redshift in respect to SUGRA.

The peaks in the lensing signal have slightly different positions in the
different models. As argued by other authors (Torri et al 2004, Meneghetti et al. 2005) 
this could be due to time offset between merger events in different 
dark energy cosmologies.

\section{Discussion and conclusions}

Models with dynamical DE are in an infant state. We do not know the
nature of DE. Thus, the  state parameter $w(t)$ is still uncertain.
In view of this functional indetermination,
at first sight, it could seem that the situation is hopeless.  

In spite of that, we can outline some general trends that
result from our analysis: in dynamical DE models, halos tend to collapse
earlier than in a \LCDM~ model with the same normalization at $z=0$.
As the result, halos are more concentrated and denser in their inner parts (KMMB03). 
Starting from this finding we have explored
the consequences of this higher concentration, on strong lensing properties of 
dark matter halos, in SUGRA and RP cosmologies.

We found that RP$_1$ halos (obtained assuming the cluster abundance of the power spectrum
and a value for the energy scale $\Lambda$ in the range suggested by 
the physics of fundamental interactions) produce a higher number of arcs 
with a $L/W>10$ if compared to the standard \LCDM~ model. This model (RP$_1$)
is marginal consistent with observations and its purpose is mainly to illustrate
the principal effect of a dynamical dark energy component on arcs statistic.

The second model we analyzed based on RP potential (RP$_2$) is more realistic
from an observational point of view but less motivated by theoretical
arguments. This model produce about 50\% more arcs with $L/W>10$ than the
\LCDM~ one for $z_l=0.3$ and $z_s=1$ but it is marginally distinguishable from \LCDM~ for
lensing system at moderate high redshift ($z_l>0.35$, fig
\ref{fig:arcsz}) or for high redshift sources/arcs ($z_l=0.3$ and $z_s=2$).

The SUGRA model is always in between the \LCDM~ and the RP$_1$ models and it
produces about 70-80\% more arcs than \LCDM~. This difference is almost
constant both changing the sources and the lens redshift and it tends to
disappear for $z_l>0.6$ (for $z_s=1.0$) where all the lensing systems considered in this
paper ($M_l \approx 5 \times 10^{14}$ ) are unable to produce highly distorted
images (except in the test model RP$_1$). We also noted that part of the stronger
lensing signal due to the higher concentration of halos in dynamical DE models
is partially canceled by geometrical effects that increase the critical surface
density in such models (fig. \ref{fig:sigma} and fig. \ref{fig:arcs2}).


As final remark we would like to stress that arc statistic is a powerful tool 
to investigate the nature of the Dark Energy.
The forthcoming observational surveys(i.e. CFHT Legacy Survey, SDSS and others) 
will improve the statistic of giant arcs on the sky (for example the RCS-2 Survey
(Gladders et al. 2003) will cover an area of 830 deg$^2$ and is expected 
to produce 50-100 new arcs).
Such an observational material will provide a discrimination between
DE cosmologies possibly allowing to constrain the $\Lambda$ scale
of the SUGRA and RP potentials.


\section*{ACKNOWLEDGMENTS}

It's a pleasure to thank Massimo Meneghetti for his help and his comments on 
lensing simulations and Roberto Mainini for useful discussion on dynamical
dark energy models. The helpful comments of an anonymous referee led to
substantial improvements in the paper. We also thank S. Bonometto and B. Moore 
for carefully reading the manuscript and INAF for allowong us to use of
the computer resources at the CINECA consortium 
(grant cnami44a on the SGI Origin 3800 machine).

\label{lastpage}

\end{document}